\documentclass[usenatbib]{mn2e}

\usepackage{graphicx}
\usepackage{multirow}
\usepackage{natbib}
\usepackage{lineno}
\def\citeN{\citet}
\def\cite{\citep}

\footnotesize
\newdimen\digitwidth    
\setbox0=\hbox{\rm0}
\digitwidth=\wd0
\catcode`!=\active
\def!{\kern\digitwidth}
\normalsize
\title[A connection between state changing and glitches]{A connection between radio state changing and glitch activity in PSR J0742$-$2822}
\author[M.~J.~Keith et al.]
{M.~J.~Keith$^{1}$\thanks{Email: mkeith@pulsarastronomy.net},
R.~M.~Shannon$^{1}$ and
S.~Johnston$^{1}$
\\
$^1$ CSIRO Astronomy \& Space Science, Australia Telescope National Facility, P.O. Box 76, Epping, NSW 1710, Australia\\
}

\let\leqslant=\leq 
\date{}
\begin{document}

\maketitle
\newcommand{\setthebls}{
}

\setthebls

\begin{abstract} 
PSR J0742$-$2822 exhibits two distinct emission states that are identified by discrete changes in the observed pulse profile.
These changes have previously been shown to correlate with changes in the derivative of the pulse frequency.
In this paper, we use observations with the Parkes radio telescope at a centre frequency of 1369 and 3100 MHz to produce high phase resolution polarisation profiles for the two modes and perform a detailed study of the correlation between observed pulse shape and spin-down rate.
We find no correlation for at least 200 days prior to a glitch in the pulsar at MJD 55022, following which the correlation becomes strong.
This suggests a link between the emission state switching phenomenon and glitch events. We discuss the possibility that emission state switching is driven by the interaction between the magnetosphere and the interior of the neutron star.

\end{abstract}

\begin{keywords}
pulsars: general ---
pulsars: individual: (PSR J0742$-$2822)
\end{keywords}

\section{Introduction}
Radio pulsars are renowned for highly regular pulsations driven by their stable rotation.
Although some pulsars have precision of tens of nanoseconds over many years \cite{mhb+13}, this is not the case for the majority of pulsars, especially those with characteristic ages less than $\sim 100$~kyr.

There are two main drivers of instability in pulsars, ``glitches'' (e.g. \citealp{lsg00,apas84}) and ``timing noise'' (e.g. \citealp{ch80,hlk10}).
Glitches are characterised by a sudden change in the observed pulse period.
These sudden changes are typically followed by a slower ``recovery'', where the period returns towards the value it would have had without the glitch.
This recovery occurs over a wide range of time-scales, although it is at present not well understood \cite{vm10}.

Timing noise describes the unmodelled red noise in the observed pulse arrival times of the majority of radio pulsars~\cite{cd85,hlk10,sc10}.
The physical processes behind both of these observationally defined phenomena are not well understood, however there are two main classes of theory. Either the variation in pulse arrival times is driven by the interior of the neutron star, or due to a change in the magnetospheric arrangement.

Most accepted theories for glitches model their origin in the interior of the neutron star, where angular momentum is stored and then suddenly transferred to the crust in order to cause a sudden change in the spin of the neutron star surface \cite{ai75,apas84}.
Timing noise is less well understood; however recent work by \citeN{lhk+10} demonstrated that for six pulsars, the observed timing noise (characterised by fluctuations in period derivative) correlates with the observed pulse shape (characterised by pulse width or relative amplitude of pulse profile components).
This suggests that some timing noise may be driven by some magnetospheric process, possibly with discrete state switching on timescales from $100$ days to greater than $2000$ days.
Similar profile changes have also been observed in PSR J0738$-$4042, with the profile shape varying on timescales of decades~\cite{krj+11}.
It has been suggested that the timing noise in the youngest pulsars is dominated by unmodelled recovery from glitches and therefore by the neutron star interior \cite{jg99}, but as the pulsar ages, glitch activity decreases and the timing noise is dominated by changes in the magnetosphere.
However, there are some indications that glitches can also be linked to pulse shape changes.
A recent glitch in PSR J1119$-$6227 coincided with the appearance of additional pulse components with intermittent or RRAT-like behaviour \cite{wje11}.

In this paper, we present a study of PSR J0742$-$2822, one of the six pulsars studied by \citeN{lhk+10}.
This pulsar exhibits the most rapid state changes of the sample, and has the least clear correlation between pulse shape changes and spin-down changes.
In addition, this pulsar has several published glitches, with the most recent (MJD 55022) being of moderate size \cite{elsk11,ymh+12}.

\section{Observations}
\label{observations}
Since 2007, PSR J0742$-$2822 has been observed on a roughly monthly basis at 1369~MHz and six-monthly at 3100~MHz with the Parkes radio telescope as part of the {\it Fermi} timing programme\cite{wjm+10}.
The data were recorded with one of the Parkes Digital Filterbank systems (PDFB1/2/3/4), with a total bandwidth of 256~MHz at 1369~MHz and 1024~MHz at 3100~MHz.
We supplemented these observations with archival data from the Parkes pulsar data archive \cite{hmm+11}.

Flux densities have been calibrated by comparison to the continuum radio source 3C\,218.
Observations are calibrated for differential gain and phase between the feeds using measurements of a noise diode coupled to the receptors in the feed.
To correct for cross-coupling of the receptors in the Multibeam receiver used for the 1369~MHz observations, we used a model of the Jones matrix for the receiver computed by observation of the bright pulsar PSR J0437$-$4715 over the entire range of hour angles visible, using the `measurement equation modelling' technique \cite{van04c}.

\section{Timing}
\label{spin-down}
We follow the same general pulsar timing procedure as used by \citeN{wjm+10} to obtain a best-fit pulsar ephemeris.
Because of the large amount of timing noise, we fit the timing model using the pre-whitening technique of \citeN{chc+11} to perform an unbiased least-squares fit in the presence of red noise.
This model includes pulse frequency, frequency derivative, position, proper motion and the previously known glitch event at MJD 55022 \cite{elsk11,ymh+12}, however it does not attempt to model the timing noise.
We measure the step change in pulse frequency associated with the glitch to be $\Delta \nu = 0.61\,\mu$Hz, consistent with that previously reported \cite{elsk11,ymh+12}.
The difference between our best-fit model and the observed time of arrival is termed the timing residual, and in the case of PSR J0742$-$2822 is dominated by a combination of red timing noise and glitch recovery.

\begin{figure}
\begin{center}
\includegraphics[width=7.5cm]{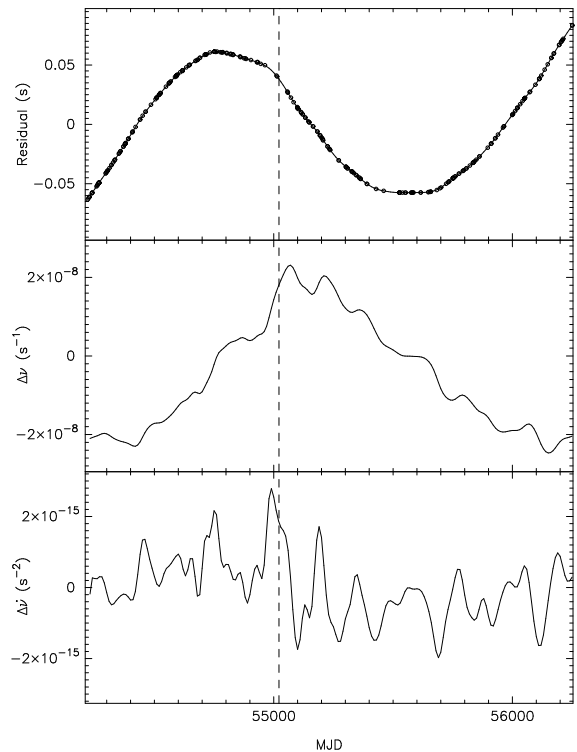}
\end{center}
\caption{\label{timing} Timing noise in PSR J0742$-$2822. Upper panel shows the post-fit timing residual for PSR J0742$-$2822. Dots show individual observations (with error bars too small to show), to which we fit an optimally constrained smooth curve (solid line). These residuals have been corrected for the discrete change in pulse frequency due to a glitch at MJD 55022, marked with a vertical dashed line.
The middle panel shows the unmodelled variation in pulse frequency, derived from the derivative of the smooth fit to the residuals.
The lower panel shows the derivative of the above, the unmodelled variation in pulse frequency derivative.  }
\end{figure}

The pulsar observations are irregularly sampled and so we fit a smooth curve to the timing residuals using the interpolation technique of \citeN{dch+12}, which gives us an effective time resolution of 60 days.
Both the residuals and the smooth fit are shown in Figure \ref{timing}.
When the slope of the residuals is positive the pulse frequency is overestimated.
By taking the gradient of the residuals, we can determine the anomalous pulse frequency, i.e. the fluctuations which are not modelled by our ephemeris.
Similarly, the second derivative gives us the anomalous frequency derivative.
These are also shown in Figure \ref{timing}.

We confirm that the frequency derivative shows quasi-periodic oscillations with a timescale on the order 100 days \cite{lhk+10}.
We also note that the mean frequency derivative after the glitch is reduced by about $-1\times10^{-15}\,$s$^{-2}$, which we attribute to the step change in $\dot\nu$ caused by the glitch.
This value of $\Delta \dot\nu$ is consistent with that reported by \citeN{ymh+12}, rather than the larger value reported by \citeN{elsk11}.
This step in $\Delta \dot\nu$ has not been removed from the Figure \ref{timing} but we correct for it in the remainder of this work.

\section{Pulse Profiles}
\label{profiles}

\begin{figure}
\begin{center}
\includegraphics[width=6.9cm]{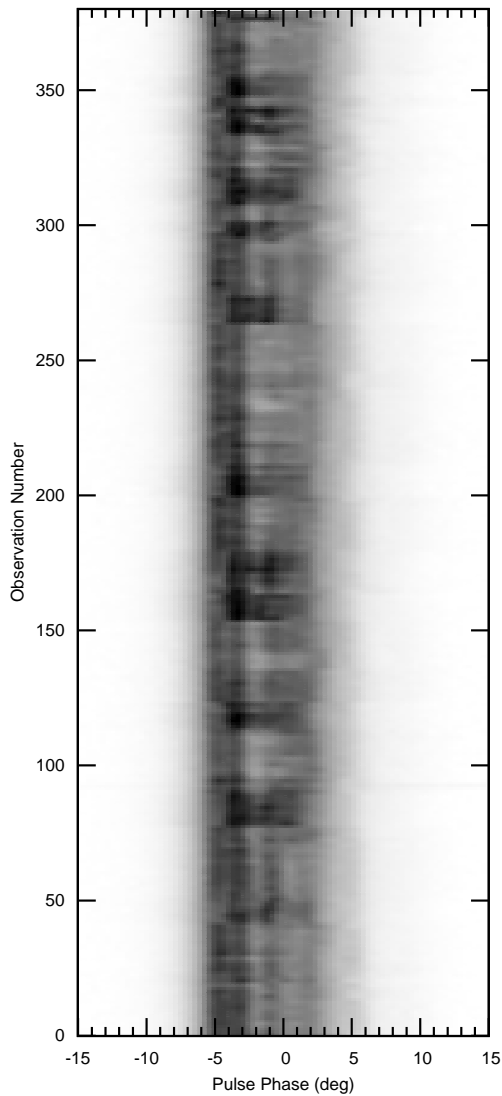}
\end{center}
\caption{\label{waterfall}Stacked profiles of 380 observations of PSR J0742$-$2822 at a centre frequency of 1369~MHz, normalised to keep the pulse width constant. Note that the observations are not uniformly spaced in time.}
\end{figure}

Average pulse profiles of PSR J0742$-$2822 exhibit complex structure with as many as seven overlapping components~\cite{kra94}.
The pulses are highly linearly polarised, with a small amount of negative circular and  slight depolarisation towards the trailing edge of the pulse~\cite{kj06}.
The viewing geometry is not well constrained, but it is likely to be close to orthogonal and the profile is symmetrically distributed around the magnetic pole \cite{jhv+05}.
The evolution of the profile with frequency is constant with a central `core' component surrounded by at least two `conal' rings, though it has been noted that the relative spectral index of some components appears to change significantly with frequency \cite{jkw06}.

\begin{figure}
\begin{center}
\includegraphics[width=7cm]{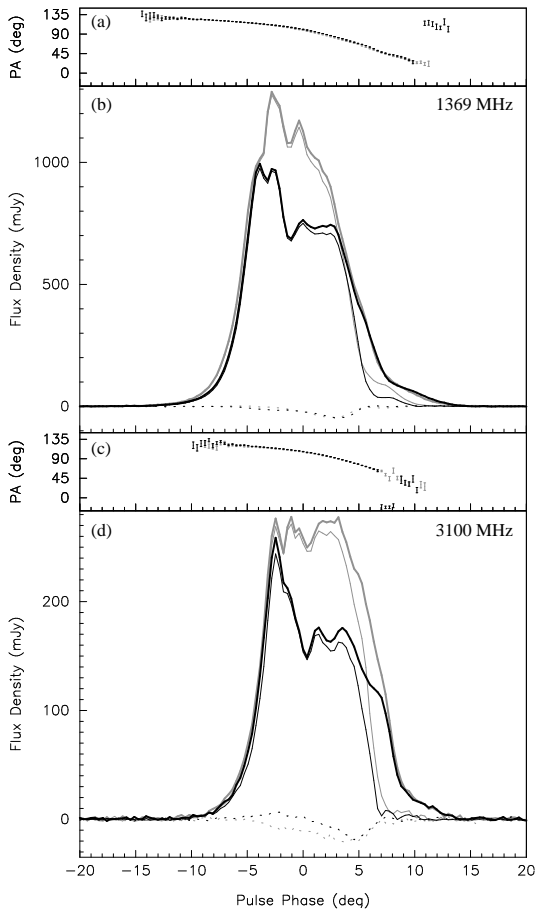}
\end{center}
\caption{
	\label{prof_20}
Average pulse profiles of PSR J0742$-$2822 for Mode I (black) and Mode II (grey).
The upper two panels, (a) and (b), are for a centre frequency of 1369~MHz, and the lower two panels, (c) and (d) are for a centre frequency of 3100~MHz.
Panels (a) and (c) show the polarisation position angle as a function of pulse phase.
In panels (b) and (d), total intensity is shown with a thick line, linear polarisation with a thin line and circular polarisation with a dotted line.
Note that for much of the profile the pulse is almost 100\% linearly polarised and so the line showing linear polarised intensity is obscured by the total intensity.
}
\end{figure}

Figure \ref{waterfall} shows a large number of integrated observations of the pulsar at a centre frequency of 1369~MHz, taken over five years, stacked sequentially.
These profiles are averages of typically three minutes of observation and have been normalised by the leading and trailing edges to account for variation in overall flux density.
Two emission modes are clearly identifiable, with the most significant change occurring between phase $\sim -4\degr$ to $-1\degr$ on this figure.
We define the most frequently observed mode to be ``Mode I'' and the less frequently observed mode to be ``Mode II''.
The two modes are not completely distinct and some profiles have an intermediate shape showing characteristics of both modes.
Nevertheless, we can split the profiles into the two modes and compare the high-resolution average profiles.
To classify the profiles we define a shape parameter based on the 1369-MHz pulse profile to be the ratio of the first and second components in the profile (the two leading peaks in Mode I; see Figure \ref{prof_20}).
For Mode I, this parameter is roughly unity, and for Mode II this is less than unity.
We find that this method is supported by subjective assessment of individual profiles, and although other parameters can be derived we find that the choice of parameter does not affect the subsequent analysis.

The upper two panels of Figure \ref{prof_20} show the average profiles for the two modes of PSR J0742$-$2822 at a centre frequency of 1369~MHz.
Here we have averaged a total of $\sim 15000\,$s for Mode I and $\sim 8000\,$s for Mode II which renders day-to-day flux density variations caused by interstellar scintillation negligible, and so normalisation is not required.
The two modes have a very similar overall pulse width, and are very similar on both the leading and trailing edge of the profile.
Mode II exhibits significantly more emission in the central part of the profile, with prominent increases in the second and third peaks of the profile as well as a general increase of the trailing edge.
Both modes retain a high linear polarisation fraction, however Mode II is slightly more polarised on the trailing edge.
The position angles are also consistent within their measurement uncertainties, except for an orthogonal mode jump observed in Mode I at the very trailing edge of the profile.

An identical analysis was performed for the observations centred at 3100~MHz, and the results are shown in the lower two panels of Figure \ref{prof_20}.
The change between the two modes is even more pronounced at the higher frequency, with the Mode I profile dominated by a leading narrow component.
The Mode II profile is almost rectangular, with the trailing half of the profile almost doubling in flux density over Mode I.
The Mode I profile again shows an orthogonal polarisation mode jump at the trailing edge of the profile, though over a different range of pulse phase to that seen at 1369~MHz.
Although not surprising, we note that the Mode I and II profiles are consistent between the two frequencies, i.e. when the pulsar is in a particular mode at one frequency, it is in that same mode at the other frequency.
The leading component, which dominates Mode I, does not appear to change amplitude significantly between the modes and has a flatter spectral index than the components that dominate in Mode II.
We note that the increased amplitude of the pulse-shape changes at high frequency may have led \citeN{jkw06} to confuse emission mode changes with frequency evolution in their multi-frequency study.

\begin{figure}
\begin{center}
\includegraphics[width=7.5cm]{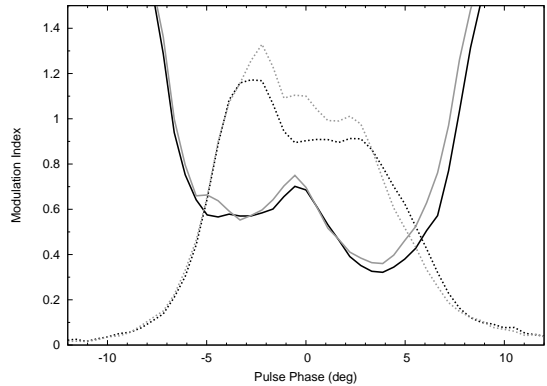}
\end{center}
\caption{\label{single} Two observations of PSR J0742$-$2822 recording 1000 individual pulses, taken on 2012-04-15 (Mode I; black lines) and 2012-05-27 (Mode II; grey lines).
The solid lines show the modulation index and dotted lines show the average profile from that observation for reference.
}
\end{figure}

A number of observations were recorded with sufficient time resolution to study individual pulses.
Figure \ref{single} shows the phase resolved modulation index (ratio of variance to mean of individual pulses) for representative observations in each of the two modes.
Due to a hardware limitation when recording individual pulses, the phase resolution of this data is half that in Figure \ref{prof_20}.
Even though the observations show changes in the mean profile corresponding to the two modes, the underlying statistics of the individual pulses remain constant and in agreement with previously published values \cite{wes06}.
Both leading components have similar modulation indices, even though one of these components varies considerably between modes and the other is almost unchanged.
The central peak seen in the modulation index is associated with a narrow component that is seen to vary considerably between observations but does not appear to be strongly correlated with the overall mode changing or timing analysis.

\section{Analysis and discussion}
\begin{figure*}
\begin{center}
\includegraphics[width=18cm]{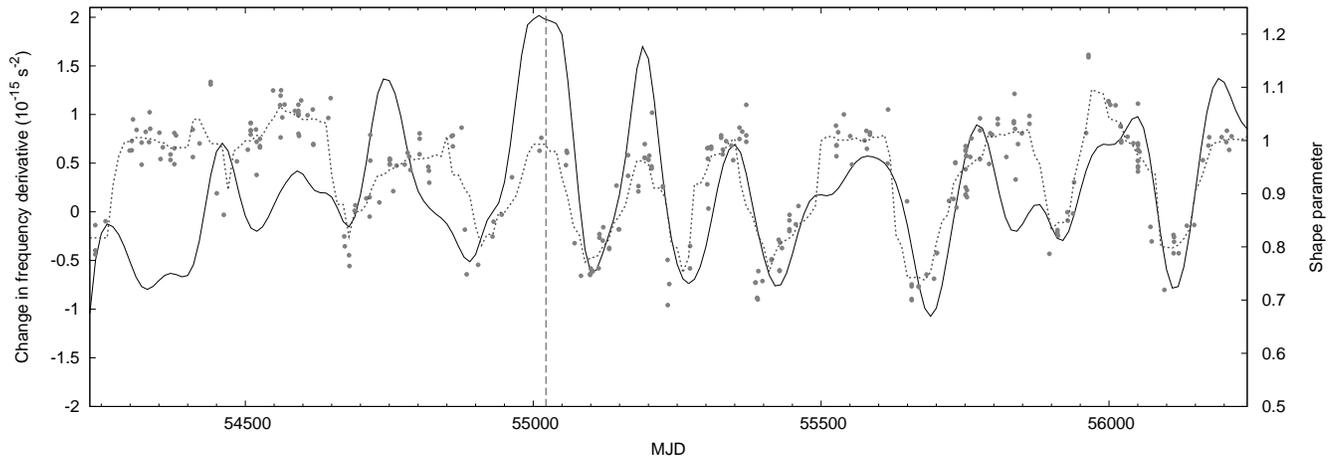}
\end{center}
\caption{
	\label{f1_prof}
Fluctuations in frequency derivative of PSR J0742$-$2822 in excess of the best-fit timing model (See Section \ref{spin-down}).
Circles indicate the measured pulse shape parameter for each of the observations.
A grey dotted line shows the shape parameters averaged under a running box of width 60 days.
The vertical dashed line at MJD 55022 indicates the epoch of a glitch.
}
\end{figure*}

In Figure \ref{f1_prof} we overlay the pulse shape parameter defined in Section \ref{profiles}, computed for each observation, with the observed frequency derivative deviations described in Section \ref{spin-down}.
We also show the pulse shape parameter averaged under a running 60-day window to match the time resolution of the $\dot \nu$ measurements.
The correlation between spin-down rate and pulse shape parameter is very clear for much of the timespan.
We find that our Mode II is associated with increased $|\dot\nu|$, i.e. larger negative $\dot \nu$.
This association of the mode dominated by steep spectral index components with a larger spin-down rate supports the hypothesis that so-called central `core' components tend to dominate the large spin-down rate mode \cite{lhk+10}.

\begin{figure}
\begin{center}
\includegraphics[width=8cm]{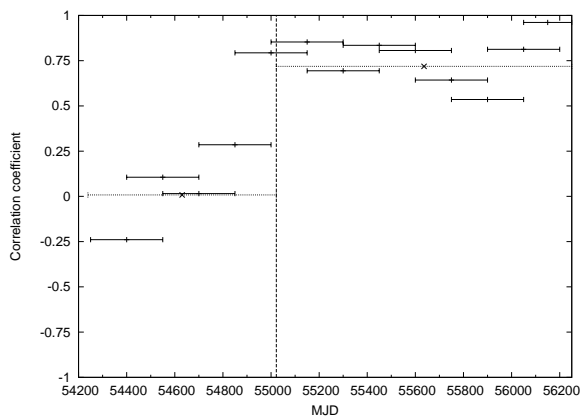}
\end{center}
\caption{
	\label{corr}
Correlation between frequency derivative and smoothed pulse shape parameter for overlapping 300-day intervals.
The vertical dashed line at MJD 55022 indicates the epoch of a glitch.
Also shown with dotted bars is the same correlation when computed for the entire pre and post-glitch epochs.
}
\end{figure}

For two years prior to the MJD 55022 glitch of PSR J0742$-$2822 the pulse profile exhibits two distinct emission states, however this does not seem to correlate strongly with the pulsar spin down parameter.
Figure \ref{corr} shows the correlation between the pulse shape parameter and $\Delta\dot\nu$ for overlapping 300-day windows.
The correlation is computed after averaging the shape parameter measurements under a running box-car of width 60 days to match the effective time resolution of the $\Delta\dot\nu$ measurements.
This confirms that the correlation swiftly increases at the glitch event of MJD 55022 and remains high for more than 1000 days.
Although we observe a change in correlation, it can also be possible that the rate of state switching is too rapid prior to the glitch for us to detect with our 60-day time resolution.
Indeed \citeN{lhk+10} show that the mode-switching rate of PSR J0742$-$2822 is the most rapid of the sample and that this rate is not constant over the entire data-span.
As we only have data for one glitch occurrence this change in correlation co-efficient may be coincidental, but we feel that it is worthwhile to consider the possibility that the change in the state-switching behaviour is linked to the glitch event.

Glitch events are generally thought to be driven by the neutron star interior, but the emission and spin-down is thought to be driven by the magnetosphere.
There are suggestions that a link between pulse shapes and glitches is present in PSR J1119$-$6227, where an unusual glitch appeared to trigger the appearance of additional components in the profile \cite{wje11}.
However, unlike PSR J1119$-$6227, we do not see any evidence that the change in emission in PSR J0742$-$2822 is associated with RRAT-like statistics of individual pulses.
We still do not have any clear picture of how the magnetospheric changes could influence or could be influenced by glitch events, nor how these might relate to the quasi-periodic spin-down changes observed in the \citeN{lhk+10} sample.

Both emission state-changing and glitch activity are associated with a wide range of timescales which, for each phenomenon, are broadly consistent across discrete events in an individual object.
For example, the time between glitches has a characteristic timescale for each pulsar, and in a few cases exhibits quasi-periodic behaviour \cite{mpw08}.
The time spent in each emission state also shows quasi-periodic fluctuations.
Glitch timescales can be used to probe the physics of the neutron star interior (e.g. \citealp{vm10}).
If glitches and emission state switching are be related then perhaps some of these timescales are driven by the same physical processes.

In the MJD 55022 glitch of PSR J0742$-$2822, the spin period of the pulsar increased by $\Delta \nu = 0.61\,\mu$Hz, which can be considered to be a lower bound on the differential rotation frequency between the neutron star crust and its interior.
In this case the value of $\Delta \nu$ implies a differential rotation period between the neutron star crust and interior of less than $\sim 19\,$days.
This is about a factor of 5 smaller than the observed `typical' periodicity in the state changes, however this differential rotation period is valid only prior to the glitch and we are only sensitive to state changes in excess of $\sim 60$ days.
We do not know what the post-glitch differential rotation period is, but it should be much longer because angular momentum has been transferred from the interior to the slower-moving crust.
Is it possible then that this differential rotation is driving the periodicity in the observed $\dot\nu$ and profile shape changes?
We note that PSR J0742$-$2822 has the shortest quasi-periodicity of all the state-changing pulsars, and exhibits moderate glitches, implying a moderate differential rotation rate.
The other pulsars have longer mode changing timescales, which we could model as a smaller differential rotation rate and therefore may not be expected to glitch as rapidly.
However, there is currently little direct evidence for any link between the differential rotation period and the spin down rate changes.
Detailed studies of other state-changing pulsars, and high-cadence studies of other likely candidates with short period variation would be greatly valuable in testing this idea.

\section{Acknowledgements}
The Parkes radio telescope is part of the Australia Telescope which is funded by the Commonwealth of Australia for operation as a National Facility managed by CSIRO.

\bibliographystyle{mnras}
\bibliography{journals,myrefs,modrefs,psrrefs,crossrefs}

\end{document}